\documentclass[twocolumn,aps,prl,amsmath,amssymb]{revtex4}
\usepackage{graphicx}
\usepackage{dcolumn}
\usepackage{bm}
\usepackage{epstopdf}
\DeclareGraphicsExtensions{.pdf,.eps,.png,.jpg,.mps}
\begin{document}
\title{Strong Plasmonic Enhancement of Photovoltage in Graphene}
\author{T. J. Echtermeyer$^1$}
\author{L. Britnell$^2$}
\author{P. K. Jasnos$^1$}
\author{A. Lombardo$^1$}
\author{R. V. Gorbachev$^3$}
\author{A. N. Grigorenko$^2$}
\author{A. K. Geim$^3$}
\author{A. C. Ferrari$^1$}
\author{K. S. Novoselov$^2$}
\affiliation{$^1$Department of Engineering, University of Cambridge,Cambridge CB3 0FA, UK}
\affiliation{$^2$School of Physics $\&$ Astronomy, University of Manchester, Oxford Road, Manchester M13 9PL, UK}
\affiliation{$^3$Centre for Mesoscience $\&$ Nanotechnology, University of Manchester, Oxford Road, Manchester, M13 9PL, UK}
\begin{abstract}
Amongst the wide spectrum of potential applications of graphene\cite{Geim2007,Geim2009}, ranging from transistors and chemical-sensors to nanoelectromechanical devices and composites, the field of photonics and optoelectronics is believed to be one of the most promising\cite{Bonaccorso2010,Blake2008,Wang2008,Liu2011,Sun2010}. Indeed, graphene's suitability for high-speed photodetection was demonstrated in an optical communication link operating at 10 Gbit/s\cite{Mueller2010}. However, the low responsivity of graphene-based photodetectors compared to traditional III-V based ones\cite{Mueller2010} is a potential drawback. Here we show that, by combining graphene with plasmonic nanostructures, the efficiency of graphene-based photodectors can be increased by up to 20 times, due to field concentration in the area of a p-n junction. Additionally, wavelength and polarization selectivity can be achieved employing nanostructures of different geometries.
\end{abstract}
\maketitle

Graphene-based photodetectors have excellent characteristics in terms of quantum efficiency and reaction time, due to the very large room-temperature mobility and high Fermi velocity of charge carriers in this material\cite{Mueller2010,Lee2008,Xia2009}. Although the exact mechanism for light to current conversion is still debated\cite{Park2009,Xu2010}, a p-n junction is usually required to separate the photo-generated electron-hole pairs. Such junctions are often created close to the contacts, due to the difference in work-function of metal and graphene\cite{Giovanetti2008,Blake2009}. Whatever the photocurrent generation mechanism, all such devices suffer from the three following problems:(i) low light absorption of graphene (2.3\% of normal incident light\cite{Nair2008,Kuzmenko2008}; (ii) difficulty of extracting photoelectrons (only a small area of the p-n junction contributes to current generation); (iii) absence of a photocurrent for the condition of uniform flood illumination on both contacts of the device. Unless the contacts are made of different materials, the voltage/current produced at both contacts will be of opposite polarity for symmetry reasons, resulting in zero net signal\cite{Mueller2010,Lee2008,Park2009}.

One possible way of overcoming these restrictions is to utilize plasmonic nanostructures placed nearby the contacts. Incident light, absorbed by such nanostructures, can be efficiently converted into plasmonic oscillations, which lead to a dramatic enhancement of the local electric field. One might consider this process as generation of evanescent photons that exist only in the near-field region\cite{Nie1997,Schedin2010,Lee2011}. Such field enhancement, exactly in the area of the p-n junction formed in graphene, can result in a significant performance improvement of graphene-based photodetectors. The role of the plasmonic nanostructures is therefore to guide the incident electromagnetic energy directly to the region of the p-n junction. Here we demonstrate that the efficiency of such devices can be 20 times larger than traditional ones\cite{Mueller2010,Lee2008,Xia2009}.

\begin{figure*}
\centerline{\includegraphics[width=170mm]{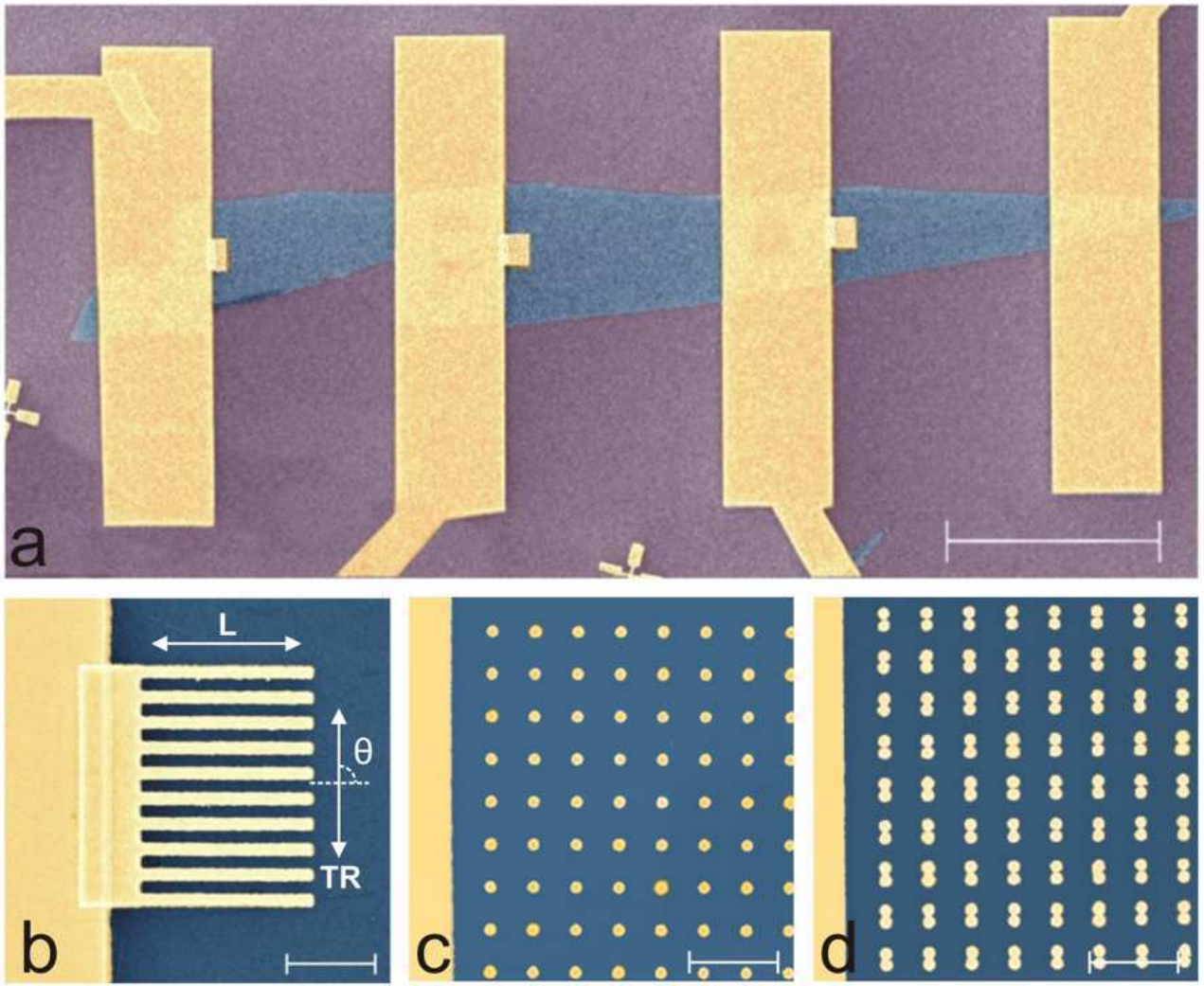}}
\caption{a) SEM micrograph of one of our samples (in artificial colors). Purple: SiO$_2$ (300nm); bluish: graphene; yellow: Ti/Au electrodes. Scale bar 20$\mu$m. b-d) Blow-up of contacts with various tested plasmonic nanostructures (in artificial colors). Longitudinal (L) and transverse (TR) incident light polarizations are indicated. Scale bars 1$\mu$m.}
\label{fig1}
\end{figure*}

We used graphene prepared by micromechanical exfoliation of graphite\cite{Grigorenko2008,Novoselov2004}. The single layer nature of our flakes was confirmed by a combination of optical contrast\cite{Blake2007,Abergel2007,Casiraghi2007}, Raman spectroscopy\cite{Ferrari2006} and Quantum Hall Effect\cite{Novoselov2005_2,Zhang2005} measurements. Ti/Au (3nm Ti, 80nm Au) contacts were formed by e-beam lithography, e-beam evaporation and lift-off. Fig.\ref{fig1}a shows the layout of the resulting devices. Various nanostructures were fabricated close to one of the macroscopic contacts of such 2-terminal devices (examples are shown in Figs.\ref{fig1}b-d). The layout and composition of the structures are chosen to produce strong light absorption in the visible range, and are similar to what we previously designed to achieve a plasmonic blackbody, resulting in almost complete absorption of incident visible light\cite{Kravets2008}. We employed several designs, but here we will mainly concentrate on one particular structure (grating with 110nm finger width; 300nm pitch Fig.\ref{fig1}b), giving the best performance.

The local photovoltage and photocurrent response of our devices is measured by coupling several lasers to a microscope, and scanning the position of the illumination spot. A Nanovoltmeter Keithley 2182A is used to record the photovoltage at the device terminals with an additional Keithley 2400 Sourcemeter, allowing control of the gate voltage. 457, 488, 514, 633 and 785nm light from multi-wavelength Ar$^+$, He-Ne and solid-state infrared lasers is coupled to the sample via a Leica DM LM microscope and a 100x ultra-long working distance objective, with a $\sim$1.5$\mu$m spot size. A PI piezoelectric stage translates the sample with respect to the laser spot in the x/y-directions, with 200nm steps, resulting in position dependent recording of the generated photovoltage. Measurements are done at room-temperature in ambient atmosphere. This allows us to measure the photovoltage dependence on intensity, wavelength and polarization of the incoming light, as well as the gate voltage. The laser power on the samples is kept$\sim$30$\mu$W. At this power, the photovoltage signal is larger than any thermopower-related signal (verified by changing the incident power). This laser intensity is also low enough not to give any observable overheating of the samples (which ensures we work in the linear regime). Raman spectra are also collected, by coupling the light scattered from the sample to a Renishaw Raman spectrometer.

\begin{figure*}
\centerline{\includegraphics[width=170mm]{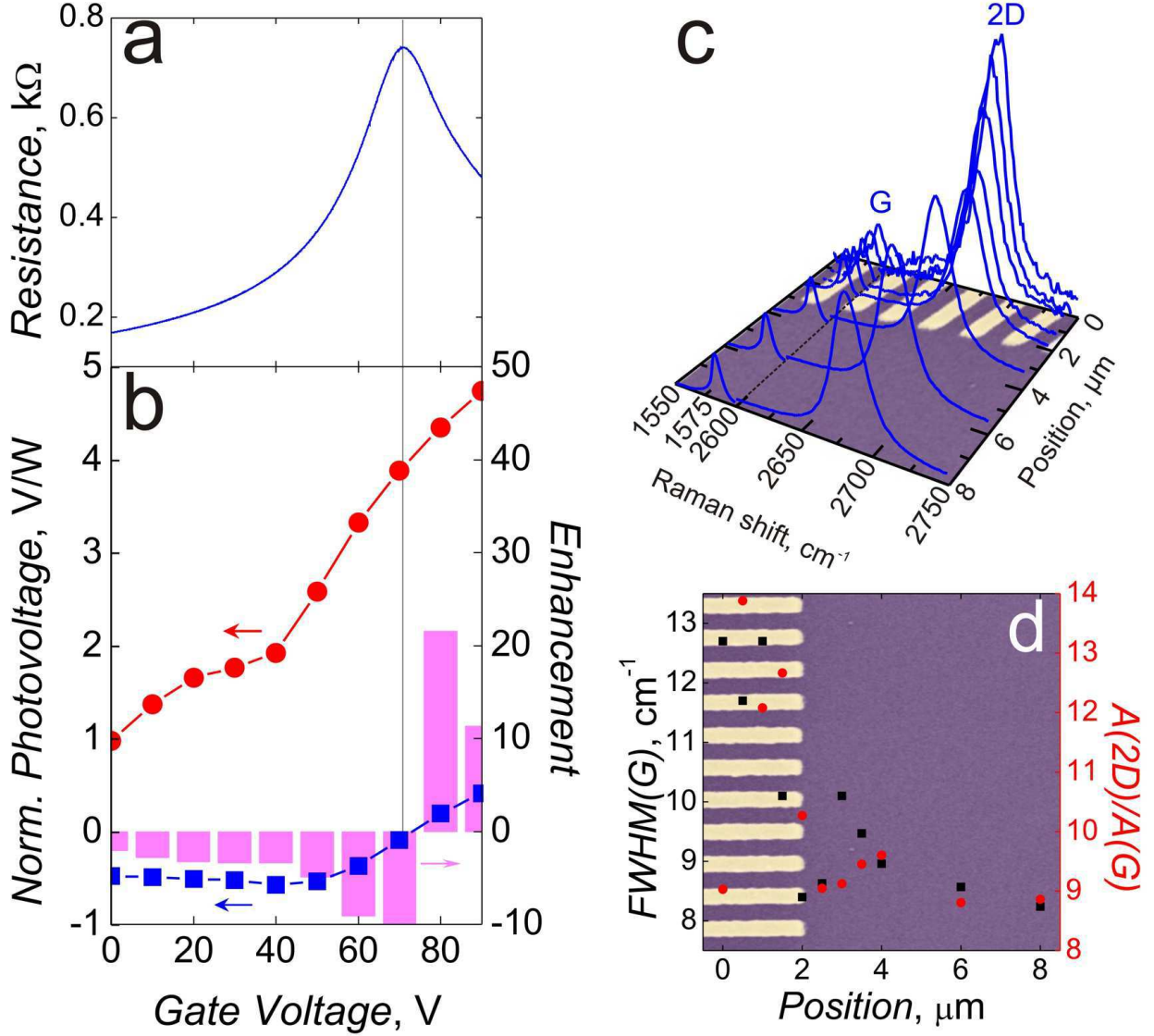}}
\caption{Resistance, photovoltage and enhancement as function of the gate voltage. a) Resistance as function of the gate voltage. b) Photovoltage for illumination close to the flat part of the contact (FC, blue), close to the structured part of the contact (SC, red) and enhancement (purple) as function of the gate voltage. Illumination wavelength: 514nm. c) Raman spectra recorded on graphene at different distances from SC. d) FWHM(G) and A(2D)/A(G) as a function of position.}
\label{fig2}
\end{figure*}

Our devices have field effect mobility$\sim$5000cm$^2$/V s at room temperature (Fig.\ref{fig2}a). They show unintentional p-doping of up to 5$\times$10$^{12}$cm$^{-2}$ (confirmed both by electronic transport\cite{Schedin2007} and Raman measurements\cite{Das2007}, (Fig.\ref{fig2}c,d), probably due to water adsorption\cite{Schedin2007}. The contacts provide local weak p-type doping\cite{Blake2009}, again confirmed by the Raman data, Fig.\ref{fig2}c,d. The photovoltage generated on the non-structured, flat part of the contact (FC), is positive for electron doping, and negative for hole-doping, as a consequence of the formation of p-n or p$^-$-p$^+$ junctions, see Fig.\ref{fig2}b.

The photovoltage generated on the structured part of the contacts (SC) is significantly higher than that on the FC. The enhancement is more than one order of magnitude for the p-n junction (Fig.\ref{fig2}b). However, the photovoltage generated on the SC has remarkably different behavior than on the FC. It is positive for all the gate voltages, monotonically decreasing for higher hole-doping, Fig.\ref{fig2}b. We do not have a complete understanding of this phenomenon, but we speculate that the most probable reason is the complex distribution of the optical electric field around the SC, allowing us to probe different parts of the p-n or p$^-$-p$^+$ junctions (which also have very complex shapes due to simultaneous screening and doping by the metal contacts) in comparison with the FC.

The doping profile is confirmed by a Raman line-scan across the contacts, carried out at zero gate voltage, Fig.\ref{fig2}c. Fig.\ref{fig2}d plots the ratios of the areas of 2D and G peaks, A(2D)/A(G), and the full width at half maximum of the G peak, FWHM(G). Far away from the contacts the Raman parameters correspond to$\sim$5$\times$10$^{12}$cm$^{-2}$ p-doping\cite{Das2007,Basko2009} (Fig.\ref{fig2}d), consistent with the transport gate-voltage measurements (Fig.\ref{fig2}a). A(2D)/A(G) significantly increases when moving close to the contacts, accompanied by a FWHM(G) increase. This implies the sample becomes less p-doped, with the area around contacts being only lightly p-doped, up to about few 10$^{11}$cm$^{-2}$\cite{Das2007,Basko2009}. In the vicinity of the SC, Fig.\ref{fig2}d shows that both FWHM(G) and A(2D)/A(G) exhibit a non-monotonous behavior, resulting in local maxima. This can be explained by the interplay between the inhomogeneous doping and strong amplification of the Raman signal around metallic nanostructures\cite{Schedin2010}.
\begin{figure}
\centerline{\includegraphics[width=90mm]{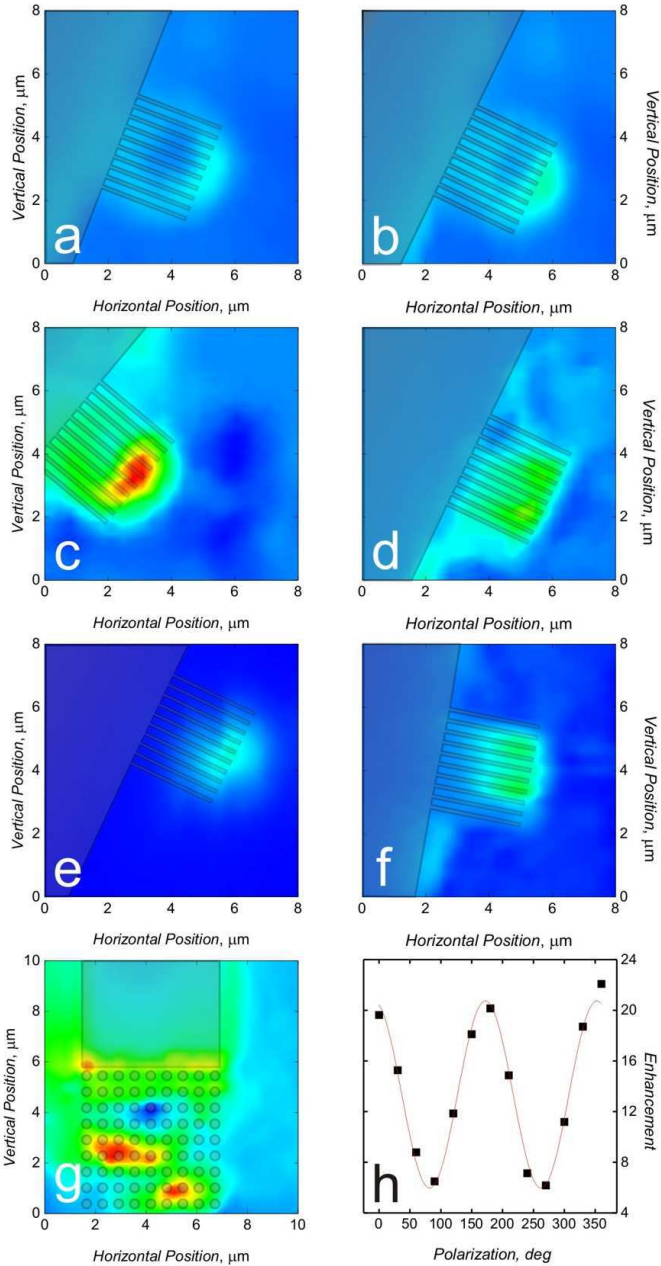}}
\caption{Photovoltage (normalized to laser power) measured on one of our nanostructured contacts (finger structure; finger width 110nm, pitch 300nm, except for g) as a function of the position of the illumination spot (spot size$\sim$1.5$\mu$m, illumination via microscope x100 objective) for various excitation wavelengths. Gate voltage 90V. Scale (except for g): from 0V (blue) to 20V (red). Overlaid is a schematic position of the contact. a) 457nm, TR polarization. b) 488nm, TR polarization. c) 514nm, TR polarization. d) 633nm, TR polarization. e) 785nm, TR polarization. f) 514nm, L polarization. g) Example of photovoltage measured on a sample with an array of nanodots. 633nm, TR polarization. Scale: from -4V (blue) to 12V (red). h) Polarization dependent enhancement at 514nm with 0$^\circ$ being TR polarization. Black squares: measured data; Red line: $\cos^2\theta$ fit.}
\label{fig3}
\end{figure}
\begin{figure}
\centerline{\includegraphics[width=90mm]{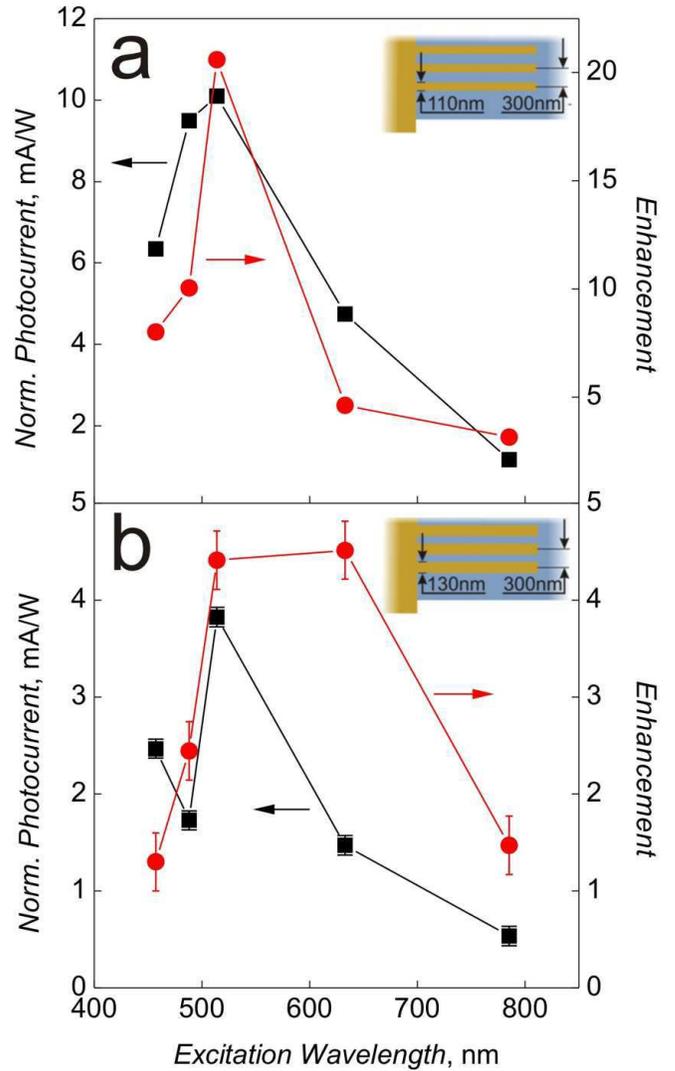}}
\caption{Photocurrent and maximum enhancement coefficient as a function of excitation wavelength for two of our finger structures of 300nm pitch. a) finger width 110 nm; b) 130nm. Insets: schematic representations of such structures.}
\label{fig4}
\end{figure}

To demonstrate the plasmonic nature of the enhancement, we mapped the photovoltaic response for different polarizations and excitation wavelengths for normal light incidence (Fig. \ref{fig3}). This allowed us to directly compare the signal produced when shining light on the FC and SC. Wavelengths covering the visible to near-infrared range (457, 488, 514, 633, 785nm, corresponding to Figs. \ref{fig3}a,b,c,d,e,f, respectively) were used. Fig. \ref{fig3} shows that the SC provides some level of enhancement for all wavelengths used (the photovoltage on SC is always larger than that on FC). The generated photovoltage is usually maximum when the laser beam is positioned at the tips of the nanostructures. This is because, in this area, both large electron band bending (due to doping from the contacts\cite{Giovanetti2008,Blake2009}) and strong enhancement of the optical field\cite{Nie1997,Schedin2010,Lee2011,Grigorenko2008} are achieved. In-between the metal stripes, although the optical field enhancement is still produced, the band bending is significantly smaller due to screening by the metal contacts.
\begin{figure*}
\centerline{\includegraphics[width=170mm]{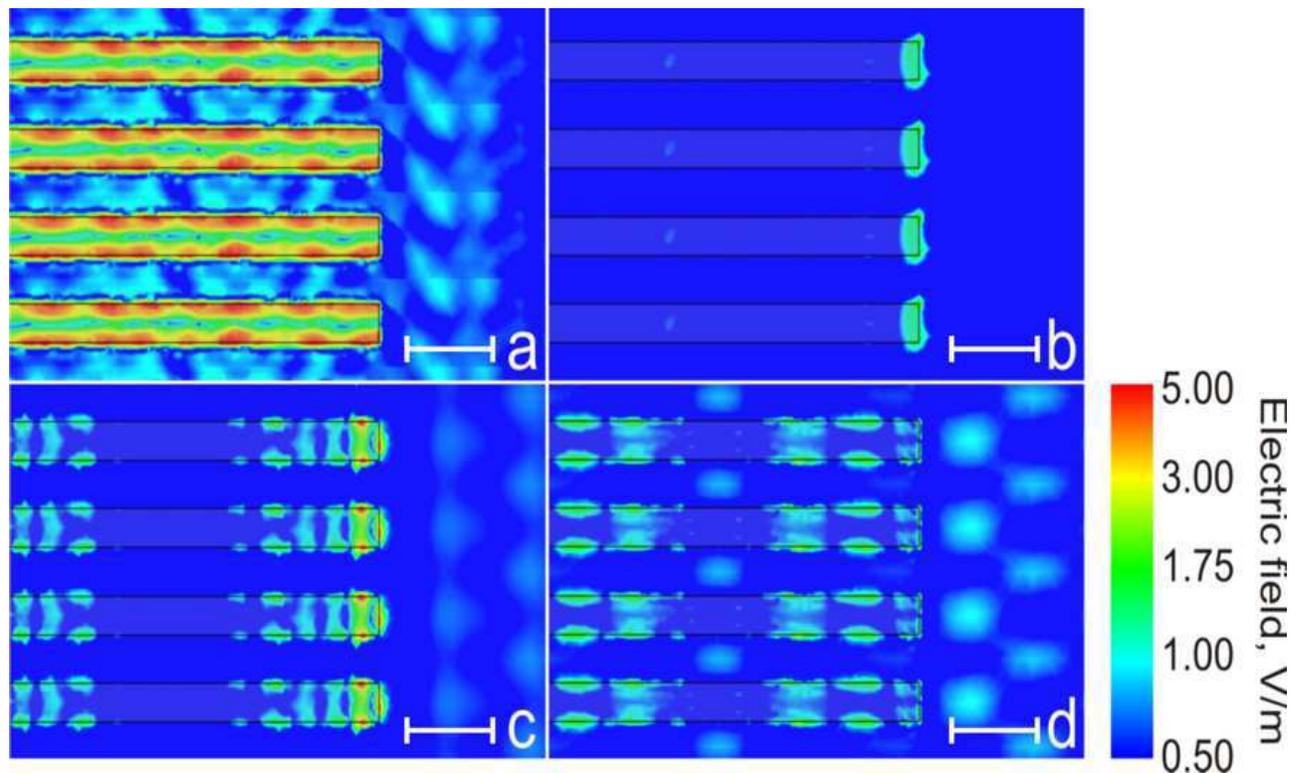}}
\caption{Numerical finite difference time domain simulations for different excitation wavelengths and polarizations. a) 514 nm, TR polarization. b) 514 nm, L polarization. c) 633 nm, TR polarization. d) 633 nm, L polarization. Scale bar 300 nm.}
\label{fig5}
\end{figure*}

We observed enhancement of the photovoltage for all wavelengths, with maximum amplification of more than 20 at the plasmonic resonance of our structure. Indeed, the strong spectral dependence of the photovoltaic enhancement suggests the importance of the plasmonic resonances in our nanostructures. The maximum enhancement for 110nm wide stripes (Figs.\ref{fig3},\ref{fig4}) is observed at 514nm (Fig.\ref{fig4}a, which is useful for solar cell applications, for instance). Depending on the SC dimensions, the resonance can be tailored to match any part of the spectrum, which might be important for applications in telecommunications. Indeed, for wider structures the resonance shifts towards larger wavelengths (e.g. the 130nm wide stripes have maximum enhancement close to 633nm, Fig.\ref{fig4}b). Such wavelength dependence rules out the possibility that this enhancement is simply due to the geometric enlargement of the junction area for the nanostructured contacts. We note that light interference in SiO$_2$ could provide some dependence of the photovoltage on the wavelength of the excited light, and can be used to enhance the signal even further\cite{Novoselov2005,Blake2007,Abergel2007,Casiraghi2007,Horng2011}. However, in our experiments the enhancement coefficient (Fig.\ref{fig4}) does not depend on the optical properties of SiO$_2$ and allows us to concentrate on the on the performance of such plasmonic nanostructures.

The SC photovoltage polarization dependence can be fitted with a $\cos^2 \theta$ function, Fig.\ref{fig3}h, where $\theta$ is the angle between the polarization and the long sides of the nanostructured "fingers" (see Fig. \ref{fig1}). The transverse (TR) polarization gives much stronger enhancement than the longitudinal (L), since the former couples resonantly to the plasmonic modes across the nanostructured fingers, matching the plasmon wavelength\cite{Maier2005}. The FC photovoltage polarization dependence is much weaker (the difference between TR and L does not exceed 30\%). We stress that, even though the far-field polarization properties of the metal stripes also shows $\cos^2 \theta$ dependence, they cannot generate any enhancement of photovoltage compared to FC. Hence the observed large anisotropy in enhancement ratio comes from the near-fields generated by plasmonic nanoresonators.

We modeled the enhancement of the electric field with the help of finite difference time-domain analysis using the High Frequency Structure Simulator (HFSS11)\cite{Kravets2010}. The actual device geometry was utilized in the model, and the optical constants of gold, graphene and the substrate were taken from Ref.\cite{Kravets2008}. Fig.\ref{fig5} shows the amplitude of the in-plane electric field around the nanostructures for incident light wavelengths of 514nm (Fig.\ref{fig5}a,b) and 633nm (Fig.\ref{fig5}b,c), and TR and L polarizations.  The results correlate well with our experimental data, see Figs.\ref{fig3},\ref{fig4}. Thus, the TR polarization for 514nm excitation (Fig \ref{fig5}a) gives very strong field enhancement on 110nm wide structures: a factor 5 in terms of field, which is a factor 25 in terms of power amplification, very similar to what we observe in our experiment, Fig \ref{fig4}. The enhancement is much weaker for 633nm excitation, again in excellent agreement with our experiments. We note, however, that one cannot draw a direct quantitative comparison between the calculated field enhancement and the measured photovoltaic signals. Indeed, the generated photovoltage depends on two factors: 1) the amplitude of the local optical field and 2) the strength and direction of the electronic band bending (built-in electric field due to the p-n junctions). The field amplification is strongly inhomogeneous, diverging near the contact edges, Fig.\ref{fig5}a. This, together with the fact that the p-n junction profile might also be non-trivial, complicates the problem. However, the qualitative correspondence between the experimental results and the theoretical predictions proves the viability of the concept of using field amplification by plasmonic nanostructures for light harvesting in graphene-based photonic devices.

In conclusions, light harvesting aided by plasmonic nanostructures helps to enhance the photovoltage signal and allows operation of such devices under flood illumination. Nanostructures with geometries resonant at desired wavelengths can be utilized in graphene-based photodetectors for selective amplification, potentially allowing light filtering and detection, as well as polarization determination in a single device at high operating frequencies. The frequency performance can be even improved in comparison with traditional devices, as the pasmonic structures add only negligible contribution to the capacitance (fractions of fF), but can significantly reduce the contact resistance. We believe that further optimization of such plasmonic nanostructures (e.g. making use of coupled or cascaded plasmon resonances\cite{Kravets2010,Kravets2008_2}) might lead to even greater photovoltage enhancement.

\section{Acknowledgements}
The authors are grateful to the Royal Society (UK), Engineering and Physical Research Council (UK), European Research Council, EU grants Rodin and Marie Curie ITN-GENIUS (PITN-GA-2010-264694), a SAIT GRO program, and Nokia Research Centre Cambridge, for financial support and to Elefterios Lidorikis for useful discussions.

\end{document}